\documentclass[prd,preprint,eqsecnum,nofootinbib,amsmath,amssymb,
               tightenlines,dvips]{revtex4}
\usepackage{graphicx}
\usepackage{bm}

\begin {document}


\def\Mrowczynski{Mr\'owczy\'nski}

\def\alphas{\alpha_{\rm s}}

\def\D{{\bm D}}
\def\L{{\bm L}}
\def\k{{\bm k}}
\def\half{{\textstyle{\frac12}}}
\def\fourth{{\textstyle{\frac14}}}

\def\p{{\bm p}}

\def\x{{\bm x}}

\def\v{{\bm v}}
\def\E{{\bm E}}
\def\B{{\bm B}}
\def\A{{\bm A}}

\def\j{{\bm j}}

\def\grad{{\bm\nabla}}

\def\tr{\operatorname{tr}}
\def\Im{\operatorname{Im}}
\def\Re{\operatorname{Re}}
\def\md{m_{\rm D}}

\def\lmax{l_{\rm max}}



\title
    {
    The Fate of Non-Abelian Plasma Instabilities in 3+1 Dimensions
    }

\author{Peter Arnold}
\affiliation
    {%
    Department of Physics,
    University of Virginia, Box 400714,
    Charlottesville, Virginia 22901, USA
    }%
\author{Guy D. Moore}
\affiliation
    {%
    Department of Physics,
    McGill University, 3600 University St.,
    Montr\'eal QC H3A 2T8, Canada
    }%
\author{Laurence G. Yaffe}
\affiliation
    {%
    Department of Physics,
    University of Washington,
    Seattle, Washington 98195--1560, USA
    }%

\date {May 24, 2005}

\begin {abstract}%
    {%
       Plasma instabilities can play a fundamental role in
       plasma equilibration.  There are similarities and
       differences between plasma instabilities
       in abelian and non-abelian gauge theories.  In particular,
       it has been an open question
       whether non-abelian self-interactions are the limiting
       factor in the growth of non-abelian plasma instabilities.
       We study this problem with 3+1 dimensional
       numerical simulations.
       We find
       that non-abelian plasma instabilities behave very differently
       from abelian ones once they grow to be non-perturbatively large,
       in contrast with earlier results of 1+1 dimensional simulations.
       In particular, they grow more slowly at late times,
       with linear rather than
       exponential dependence of magnetic energy on time.
    }%
\end {abstract}

\maketitle
\thispagestyle {empty}


\section {Introduction}
\label{sec:intro}

A fundamental problem in the theoretical study of quark-gluon plasmas
(QGPs) is
to understand how such plasmas equilibrate.  By what processes would
a heavy ion collision first produce a quark-gluon plasma that is
in approximate local equilibrium and/or expanding hydrodynamically?%
\footnote{
  Hydrodynamic behavior does not necessarily require local
  thermal equilibrium, as discussed in Ref.\ \cite{instability_prl}.
}
At the very least,
it would be useful and interesting to
answer this question in a simplifying theoretical limit: that of
arbitrarily high-energy collisions, where the running strong
coupling
$\alphas$ at relevant scales is arbitrarily small due to asymptotic
freedom.
Even this weak-coupling limit is rich and complicated.
Various authors have used weak-coupling techniques to study the initial
creation and interaction of the non-perturbatively dense
small-$x$ glue which will eventually
develop into the quark-gluon plasma
\cite{saturate1,saturate2,saturate3,saturate4,saturate5,saturate6,saturate7}.
Baier {\it et al.}\ \cite{BMSS}
have investigated the effects of two-particle scattering and Bremsstrahlung
on subsequently equilibrating the plasma, once the initial
small-$x$ glue has
expanded and diluted to perturbative densities, where it can be
treated as a collection of individual gluons.
However, it now seems clear that the process of thermalization is not
controlled solely by such individual particle
collisions in the weak-coupling limit \cite{ALM};
one must account for collective processes in the plasma in the
form of plasma instabilities.
The instabilities are known as Weibel or filamentary
instabilities \cite{weibel}, and they have a very long history in
traditional plasma physics.
Their possible relevance to QGP equilibration has been proposed for roughly
twenty years by \Mrowczynski\ and others
\cite{heinz_conf,mrow0,selikhov1,selikhov2,selikhov3,pavlenko,mrow1,mrow2,%
mrow3,mrow&thoma,randrup&mrow,strickland}.

\begin{figure}[ht]
\includegraphics[scale=0.40]{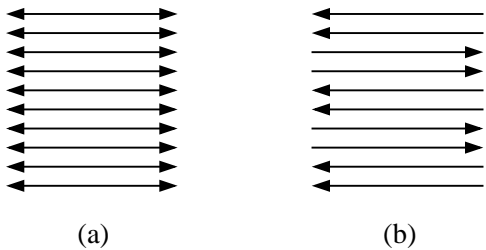}
\caption{%
    \label{fig:filaments}
    (a) Two uniform, inter-penetrating beams of charged particles,
    moving left and right; (b) the filamentation of those beams
    by the Weibel instability.
}
\end{figure}

In the case of electromagnetism, a simple example of a Weibel
instability appears for two uniform, interpenetrating beams
of charged particles traveling in opposite directions,
as depicted in Fig. \ref{fig:filaments}a.
Crudely think of each beam as a superposition of many wires carrying
current, and recall that parallel wires magnetically
attract if their currents
are aligned, or repel if opposite.
The ``wires'' are therefore unstable to clumping, as
depicted in Fig.\ \ref{fig:filaments}b.
More thorough discussions of
the qualitative origin of Weibel instabilities
may be found in Refs.\ \cite{mrow3,chen}
and \cite{ALM}.
Very roughly speaking, Weibel instabilities occur in collisionless
plasmas whenever the velocity distribution of the plasma is
anisotropic in its rest frame.
(See Ref.\ \cite{ALM} for a more precise statement.)
For the purpose of analyzing the instability,
the plasma may be regarded as collisionless whenever the distance
and time scales associated with the instability are found
to be small compared to those for individual, random collisions
of particles in the plasma.%
\footnote{
  See Ref.\ \cite{ALM} for an analysis of this point in the
  context of the bottom-up scenario \cite{BMSS} for thermalization.
}

There are magnetic fields associated with the filaments in
Fig.\ \ref{fig:filaments}b.  As the instability progresses, these
fields become stronger and stronger.  Growth of the instabilities
is initially exponential in time.  In electromagnetic
plasmas, growth stops only when the magnetic fields
become so large that they have a non-perturbatively large effect
on the particle trajectories.  These effects of these large fields
can drive isotropization of plasmas that are initially anisotropic
\cite{califano}.  This isotropization due to collective plasma
phenomena is the basis of our own recent scenario (with
Lenaghan) for the early
onset of hydrodynamic behavior in the weak coupling limit
\cite{instability_prl}, as compared to estimates based on individual
2-particle collisions.
To analyze correctly the effects of
plasma instabilities, however, it is crucial to understand whether
the (chromo-)magnetic fields grow as large in non-abelian gauge
theories as in abelian ones.  Because non-abelian fields interact
with each other, there are two possibilities for what might limit
their growth.  They might stop growing when (i) their effect on
typical particles in the plasma becomes non-perturbative, as in
the abelian case, or (ii) when their interactions with each other
become non-perturbative.  The first case occurs when
gauge fields $A$ and magnetic fields $B$ are of order \cite{AL}
\begin {equation}
   A \sim \frac{p_{\rm part}}{g}
   \qquad
   \mbox{and}
   \qquad
   B \sim \frac{k_{\rm field} \, p_{\rm part}}{g} ,
\label {eq:Ba}
\end {equation}
where $p_{\rm part}$ is the typical momentum of the particles, and
$k_{\rm field}$ is the wavenumber associated with the Weibel
instability.  The second case corresponds to
\begin {equation}
   A \sim \frac{k_{\rm field}}{g}
   \qquad
   \mbox{and}
   \qquad
   B \sim \frac{k_{\rm field}^2}{g} \,.
\label {eq:Bna}
\end {equation}
For weakly-interacting plasmas at times late enough that
the particles have diluted to perturbative densities
(number densities small compared to $p_{\rm part}^3/\alpha_s$), one finds
that
\begin {equation}
   k_{\rm field} \ll p_{\rm part} ,
\label {eq:separation}
\end {equation}
and so the second scale
(\ref{eq:Bna}) is parametrically smaller than the first
(\ref{eq:Ba}).

Based on arguments about the form of the magnetic potential
energy in anisotropic plasmas,
Arnold and Lenaghan \cite{AL} conjectured that the fields
associated with non-abelian plasma instabilities are dynamically
driven to line up in color space at the scale (\ref{eq:Bna}) when
their self-interactions become important, and that they then grow
as approximately abelian configurations to the larger scale
(\ref{eq:Ba}).  They tested this conjecture numerically in a
simplified 1+1 dimensional toy model of QCD fields in an anisotropic
plasma.  Subsequently, Rebhan, Romatschke, and Strickland \cite{RRS}
simulated the full hard-loop effective theory of the problem
in 1+1 dimensions.
They indeed found unabated exponential growth beyond the non-abelian scale
(\ref{eq:Bna}),
although the abelianization of the fields was not as global as that
of the earlier toy model simulations.

\begin{figure}[t]
\includegraphics[scale=0.80]{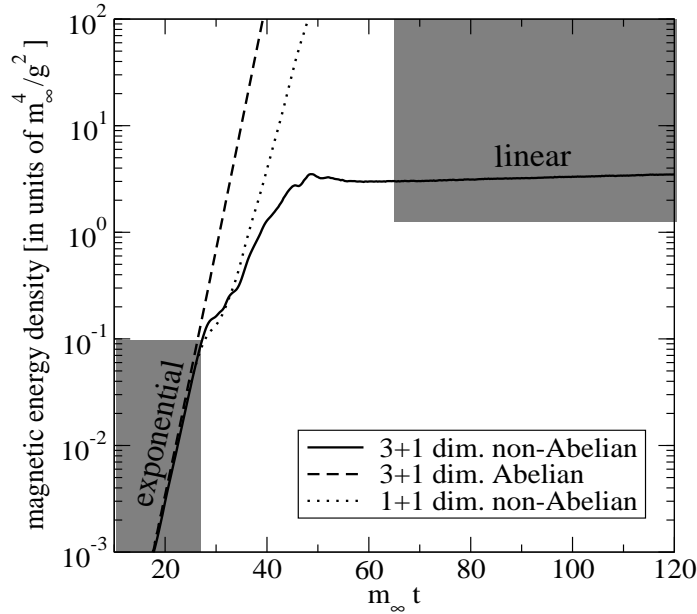}
\caption{%
    \label{fig:example}
    A representative simulation of instability growth for
    3+1 dimensional SU(2) gauge theory (solid line),
    3+1 dimensional abelian gauge theory (dashed line),
    and 1+1 dimensional SU(2) gauge theory (dotted line).
    The latter is qualitatively similar to the results of
    Ref.\ \cite{RRS}.
    The parameters used for these simulations are explained
    in Sec.\ \ref{sec:simulate}.
    For 3+1 dimensions, they are
    $\lmax = 24$,
    lattice spacing $a = 0.25 \, m_\infty^{-1}$,
    time step $0.1\,a$,
    volume $L^3 = (64 \, a)^3 = (16 \, m_\infty^{-1})^3$,
    and initial amplitude $\Delta=0.02 \, m_\infty^{-1/2}$.
    For 1+1 dimensions, they are the same except that the length is
    $L = 8192 \, a = 2048 \, m_\infty^{-1}$ and
    $\Delta=0.014 \, m_\infty^{+1/2}$.
    To simplify comparison of the curves, we have lined them up
    at early times by
    shifting the origin of time for the solid and dashed
    lines, which depend on details of initialization.
}
\end{figure}

\begin{figure}[ht]
\includegraphics[scale=0.80]{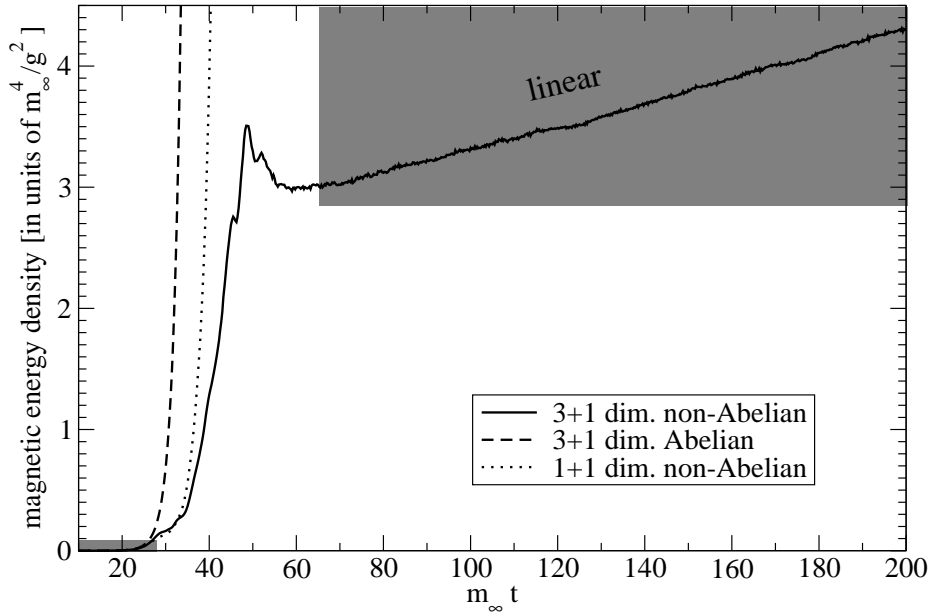}
\caption{%
    \label{fig:ex_linear}
    Same as Fig.\ \ref{fig:example}, except that the vertical
    axis is linear rather than logarithmic, and the time axis has
    been extended.
}
\end{figure}

The purpose of this paper is to investigate, through simulations,
whether the full 3+1 dimensional theory behaves similarly.  We find
significant differences.  We will soon discuss the details of
precisely what we simulate, but here we give a preview of our
results and conclusions.
The solid line in Fig.\ \ref{fig:example}
shows the growth of magnetic energy with time for a representative
simulation.  For comparison, the dashed line shows a similar
3+1 dimensional abelian simulation, and the dotted line a
1+1 dimensional non-abelian simulation.
In the lower-left shaded region of the plot, the fields are perturbative,
and all the curves grow at an exponential rate (as predicted by linearized
analysis of the instability).  For the non-abelian curves, a change
takes place in the unshaded region, which turns out to roughly
correspond to the non-abelian scale (\ref{eq:Bna}) where
self-interactions of the fields become important.  The 1+1 dimensional
non-abelian simulation soon resumes exponential growth.
The 3+1 dimensional non-abelian simulation, however, has a different
large-time behavior.  Though the magnetic energy continues to
grow with time, it eventually becomes linear with time, rather
than exponential, in the upper-right
shaded region of Fig.\ \ref{fig:example}.
This can be seen more clearly in the linear-axis plot of Fig.\
\ref{fig:ex_linear}.

As we shall explain below, the hard-loop
effective theory that we simulate
treats the plasma particles as having arbitrarily large momentum,
$p_{\rm part}\to\infty$, and so the ultimate scale (\ref{eq:Ba})
limiting growth of magnetic fields is pushed off to infinity
and irrelevant to the interpretation of our results.
These simulations are designed for the sole purpose of cleanly
studying what happens as one passes through the
non-abelian scale (\ref{eq:Bna}).
For similar reasons, the dependence of our results on $g$ is
determined by trivial scaling arguments: In the effective theory we use,
all dependence on $g$ can be absorbed by simple rescaling.
The only assumptions are that $g$ is small enough, and the separation
(\ref{eq:separation}) of scales significant enough, for the hard-loop
effective theory to be valid.
Given these assumptions, the numerical results we quote are valid for any $g$.

The primary motivation of studying instabilities in non-Abelian plasmas is to
help understand what happens in an expanding quark-gluon plasma produced
in a heavy-ion collision.  In this work, however, we simulate
non-expanding systems, both for simplicity and to isolate the particular
issue we are studying.  Ignoring the expansion is a perfectly good
approximation in cases where the instability growth rate is large
compared to the expansion rate.  For example, in the context of the original
``bottom-up'' thermalization scenario of Baier {\it et al.}\
\cite{BMSS}, this approximation would be valid \cite{ALM}
for the initial,
exponential growth of instabilities
at times late compared to the
saturation time scale $1/Q_{\rm s}$ which characterizes the initial
moments of the collision.%
\footnote{
  In this work, we will not attempt to deduce the
  ultimate effects of self-consistently
  including the physics of instabilities in the
  bottom-up scenario.  For a recent attempt at making progress in this
  direction, see Ref.\ \cite{shoshi}.
}
(The saturation momentum scale $Q_{\rm s}$ characterizes the momenta of the
original, non-perturbative, small-$x$ gluons that
eventually develop into the quark-gluon plasma in the
saturation picture.)

The rest of this paper gives details of our simulations.
In Sec.\ \ref{sec:simulate}, we explain our formulation and
discretization of the hard-loop effective theory and our choice of
initial conditions.  Sec.\ \ref{sec:more} gives further
simulation results that aid in understanding the nature of the
instability in the non-perturbative regime.
In Sec.\ \ref{sec:checks}, we discuss sources of systematic errors
in our simulations and argue that the qualitative behavior shown in
Fig.\ \ref{fig:ex_linear} is not a simulation
artifact.
Finally, we offer some last thoughts in Sec.\ \ref{sec:discussion}.
Some technical results used in the paper are left for appendices.


\section {What we simulate}
\label {sec:simulate}

\subsection {Equations of motion}

As in the 1+1 dimensional studies of Rebhan {\it et al.}\ \cite{RRS},
the starting point for our 3+1 dimensional simulations will be
hard-loop effective theory \cite{braaten&pisarski,MRS}.
This is equivalent to studying kinetic
theory of particles in the plasma, coupled to soft gauge fields,
in an approximation where the effect of the soft fields on particle
trajectories is taken to be perturbative
\cite{mrowKinetic,HeinzKinetic,BIkinetic,kelly}.
This description is valid
when there is a separation of scales $k_{\rm soft} \ll p_{\rm part}$
as in (\ref{eq:separation}), and when the soft fields have not reached the
ultimate limiting amplitude (\ref{eq:Ba}).  Excitations with momenta
of order $p_{\rm part}$ ({\it e.g.}\ the initial post-collision
gluons that provide
the starting point for the formation of the quark-gluon plasma) are
grouped together and described by a classical phase space density
$f(\p,\x,t)$.  The softer fields associated with the instability, with
momenta of order $k_{\rm soft}$, are described by classical gauge
fields.
Studying the instability by treating these fields as classical is
a good approximation because the instability quickly drives the fields to
become classically large.

Our formulation of the continuum effective theory, presented
below, is equivalent to that used
by Rebhan {\it et al}.
The theory must be discretized for simulations, and we use a
different method for discretizing velocity space than
Rebhan {\it et al.}, which we describe below.

We start with the usual kinetic theory description of a collisionless plasma
in terms of $f(\x,\p,t)$ and soft gauge fields $A_\mu(\x,t)$.
Then write $f = f_0(\p) + \delta f(\p,\x,t)$ and linearize the theory in
$\delta f$, which corresponds to a perturbative treatment of the
effect of soft fields on hard particles.  The result is well
known to have the form \cite{mrowKinetic}
\begin {subequations}
\label {eq:basic1}
\begin {equation}
    [ (D_t + \v \cdot \D_\x) \, \delta f ]_a
    +
    g \, (\E + \v \times \B)_a \cdot \grad_{\p} f_0
    =
    0 \, ,
\end {equation}
\begin {equation}
    (D_\nu \, F^{\mu\nu})_a = j^\mu_a
    \equiv
    g t_R \, \int \frac{d^3p}{(2\pi)^3} \>  v^\mu \, \delta f_a \, ,
\label {eq:basic1b}
\end {equation}
\end {subequations}
where $\delta f$ is in the adjoint color representation and
$a$ is an adjoint color index.  Here $D$ is the adjoint-representation
gauge-covariant derivative.  The group factor
$t_R$ is defined by
$\tr(T_R^a T_R^b) = t_R \delta_{ab}$, where $T_R$ is the color
generator for the color representation $R$ of the particles,
and there is an implicit sum over particle species and spins
in (\ref{eq:basic1b}).
We use $({-}{+}{+}{+})$ metric convention.

For ultra-relativistic plasmas, it is possible to integrate out the
dependence on $|\p|$ by defining%
\begin {equation}
   W_a(\v,\x) =
   g t_R \int_0^\infty \frac{4\pi p^2 \>dp}{(2\pi)^3} \>
   \delta f_a(p \v,\x,t) ,
\end {equation}
where $\v = \hat\p$ is a unit vector.
Eqs. (\ref{eq:basic1}) then imply
\begin {equation}
    (D_t + \v \cdot \D_\x) \, W
    +
    g^2 t_R  (\E + \v \times \B) \cdot 
    \int_0^\infty \frac{4\pi p^2 \>dp}{(2\pi)^3} \> \grad_{\p} f_0
    =
    0 \, ,
\label {eq:basic2a}
\end {equation}
\begin {equation}
    D_\nu \, F^{\mu\nu} = j^\mu
    =
    \int_\v v^\mu W \, ,
\label {eq:basic2b}
\end {equation}
where $\int_\v$ indicates integration over the unit sphere, normalized
so that
\begin {equation}
   \int_\v 1 = \int \frac{d\Omega_\v}{4\pi} = 1 .
\end {equation}
It turns out that these equations only depend on $f_0$ through the
angular function
\begin {equation}
   {\cal M}(\v) \equiv
   g^2 t_R \int_0^\infty \frac{4\pi p^2 \>dp}{(2\pi)^3} \>
   \frac{f_0(p\v)}{p} \,,
\end {equation}
which was introduced in Ref.\ \cite{ALM}.  Specifically, we show
in Appendix \ref{app:M} that (\ref{eq:basic2a}) can be rewritten
as
\begin {equation}
    (D_t + \v \cdot \D_\x) \, W
    + [\E\cdot(\grad_\v-2\v) - \B\cdot(\v\times\grad_\v)] {\cal M}
    =
    0 ,
\label {eq:W1}
\end {equation}
where $\grad_\v$ is the gradient operator for the two-dimensional
curved space $S^2$ of $\v$'s.  More concretely, $\grad_\v$ can be
related to the ordinary three-dimensional gradient by writing
\begin {equation}
  \nabla_\v^i = (\delta^{ij} - v^i v^j) |\p| \nabla_\p^j
\label {eq:gradv}
\end {equation}
and
\begin {equation}
   \grad_\v \> {\cal M}(\v) =
   |\p| \grad_\p \> {\cal M}\!\left(\frac{\p}{|\p|}\right)
   .
\end {equation}

Everything relevant about the initial distribution is specified
by the angular function ${\cal M}(\v)$.  This can be split into
a single dimensionful scale
\begin {equation}
   m_\infty^2 \equiv \int_\v {\cal M}
\end {equation}
and a dimensionless angular function
\begin {equation}
   \Omega(\v) \equiv \frac{{\cal M}(\v)}{m_\infty^2} .
\end {equation}
The mass $m_\infty$ turns out to be the effective mass in the dispersion
relation $\omega^2 \simeq p^2 + m_\infty^2$ for large-momentum
transverse plasmons ($p \gg m_\infty$) \cite{boltzmann}.
In the isotropic case, $\Omega=1$ and $m_\infty = \md/\sqrt2$,
where $\md$ is the Debye mass.

In order to discretize the problem for simulation,
we generalize the procedure used by B\"odeker, Moore, and Rummukainen
\cite{BMR} for the isotropic case.
Specifically, we will discretize the 2-dimensional velocity
space by expanding functions of $\v$ in spherical harmonics
$Y_{lm}(\v)$ and truncating at some maximum value $\lmax$ of $l$.
We will have to check later, of course, that our simulation results
are insensitive to the exact value of $\lmax$ used, provided
it is large enough.  So, we write
\begin {subequations}
\label {eq:lmexpand}
\begin {equation}
   W^a(\v) =
   \sum_{l \le \lmax} \sum_m
   W^a_{lm} \, \bar Y_{lm}(\v) ,
\end {equation}
\begin {equation}
   \Omega(\v) = \sum_{l \le \lmax} \sum_m \Omega_{lm} \, \bar Y_{lm}(\v) ,
\end {equation}
\end {subequations}
where we have chosen to normalize spherical harmonics without the
usual factor of $\sqrt{4 \pi}$, so that $\bar Y_{00} = 1$ and
\begin {equation}
  \int_\v \bar Y_{lm}(\v)^* \, \bar Y_{l'm'}(\v)= \delta_{ll'}
  \,\delta_{mm'}\,.
\label{eq:Ynorm}
\end {equation}
The bar over the $Y$ serves as a reminder of this normalization.
For working in $lm$-space, it is convenient to rewrite the
equation (\ref{eq:W1}) for $W$ in the equivalent form
(discussed in Appendix \ref{app:M})
\begin {equation}
    (D_t + \v \cdot \D_\x) \, W
    + \bigl(\E\cdot(\half [\v,L^2] - \v) - i \B\cdot\L\bigr) {\cal M}
    =
    0
\label {eq:W2}
\end {equation}
where
\begin {equation}
   \L \equiv -i\v \times \grad_\v = -i\p \times \grad_\p
\label {eq:L}
\end {equation}
is the operator, analogous to angular momentum, for which the
eigenvalues of $L^2$ and $L_z$,
acting on $\bar Y_{lm}(\v)$, are $l(l+1)$ and $m$.  One may
then find the coupled equations for the $W_{lm}$'s using manipulations
familiar from the quantum mechanics of spin.

In this paper, we will only simulate initial particle distributions
$f_0(\p)$ which are axially-symmetric about the $z$ axis, for which
there are many simplifications.  In this case,
\begin {equation}
   \Omega(\v) = \sum_{l \le \lmax} \Omega_{l0} \, \bar Y_{l0}(\v) .
\end {equation}
The equations for the $W_{lm}$'s in this case are given explicitly in
Appendix \ref{app:M}.  They are the same as those used by
B\"odeker {\it et al.}\ \cite{BMR} except for the $\E$ and $\B$ terms
in the $W$ equation (\ref{eq:basic2a}), which are sensitive to anisotropy in
$f_0$.  With this exception,
our discretization of the problem, and the evolution
algorithm we use, are identical to Ref.\ \cite{BMR}.
For computational simplicity, we simulate SU(2) gauge theory instead of
SU(3) gauge theory.  We do not know of any reason that SU(3) would be
qualitatively different.
For future reference, we note that
the vector current in Maxwell's equation (\ref{eq:basic2b})
is determined by the $l{=}1$ components of $W$ as
\begin {subequations}
\label {eq:j}
\begin {eqnarray}
   j_x &=& \frac{1}{\sqrt6} \, (W_{1,-1}-W_{11})
   = - \sqrt{\frac23} \Re W_{11} ,
\\
   j_y &=& - \frac{i}{\sqrt6} \, (W_{1,-1}+W_{11})
   = \sqrt{\frac23} \Im W_{11} ,
\\
   j_z &=& \frac{1}{\sqrt3} \, W_{10} .
\end {eqnarray}
\end {subequations}

We should mention that, in the traditional plasma physics literature,
the 1+1 dimensional simulations of Rebhan {\it et al.}\ would be
referred to as 1D+3V simulations, indicating that they treat 
gauge fields and particle distributions as depending on only four
of the six dimensions of phase space: one
dimension of $\x$ space and three dimensions of momentum (velocity)
space.
All three spatial components $(A_x,A_y,A_z)$ of 3-dimensional gauge
fields $\A$ are simulated in 1D+3V,
but they depend on only one spatial dimension,
{\it e.g.}\ $A_i = A_i(t,z)$.
The 3+1 dimensional simulations in this paper are correspondingly referred to
as 3D+3V.
However, in the ultra-relativistic limit, calculations
are simplified by the fact that
velocity space is effectively 2 dimensional, since $|\v|=1$.

Other 1D+3V simulations of non-abelian plasma instabilities have been
performed by Dumitru and Nara \cite{Dumitru}.
Instead of working with a phase
space distribution $f$, they simulate a finite number of discrete,
classical particles with classical color charges interacting with
the soft fields \cite{wong,christina,hu&muller,MHM}.
If one linearizes in the perturbations to the
hard particles, this formulation leads to an effective theory equivalent
to that above.  However, they do no such linearization, since their
interest lay in studying what happens if the effects on hard particles
eventually become substantial.%
\footnote{
  It is quite interesting to study the
  non-linearized theory, but a few caveats of interpretation should be
  kept in mind.  At some time, energy loss of the hard particles through
  hard Bremsstrahlung (catalyzed by interaction with the soft fields) may
  become an important process.  Processes which change the number of
  particles with momenta of order $p_{\rm part}$ cannot be described with
  the collisionless Boltzmann equation.  Further, the lattice
  implementation via ``particle and cell'' codes, as used in
  \cite{MHM,Dumitru}, suffers from spurious interactions between
  particle degrees of freedom and the most ultraviolet lattice modes,
  which may be problematic in 3D+3V simulations.  For a discussion, see
  Ref.~\cite{MHM}.
}


\subsection {Choice of $f_0$}
\label {sec:f0}

\begin{figure}[ht]
\includegraphics[scale=0.70]{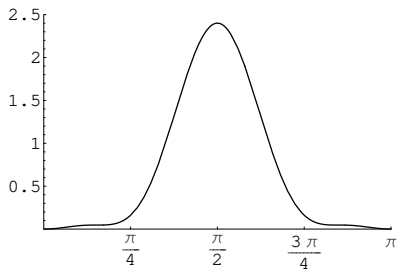}
\caption{%
    \label{fig:fchoice}
    The angular dependence $\Omega(\theta)$ of our
    hard-particle distribution (\ref{eq:fchoice}).
}
\end{figure}

If we measure all quantities in units determined by the single
dimensionful scale $m_\infty$, then the only sensitivity of our
problem to the choice of the initial hard particle distribution
$f_0(\p)$ is through the angular function $\Omega(\v)$.
In the axi-symmetric case, that means we have to choose
$\Omega(\theta)$, where $\theta$ is the angle of $\v$ with the
$z$ axis.  In this paper, we will only investigate results for
a single choice of $\Omega(\theta)$.  Our criteria for choosing
this distribution
were that (i) perturbatively, the dominant instability should be
a Weibel instability;%
\footnote{
  For a discussion, in the QCD literature, of other possibilities
  such as electric charge separation (Buneman) instabilities, see
  Refs.\ \cite{ALM,strickland}, as well as references to the traditional plasma
  literature in Ref.\ \cite{ALM}.
}
(ii) $\Omega_{lm}$ should be dominated by
relatively low $l$'s, to aid in numerical convergence of our
simulations to the $\lmax{\to}\infty$ limit; (iii) the Weibel instability
growth rate should be reasonably large (relative to wavelength),
to help minimize
simulation time; (iv) $\Omega(\theta)$ should be everywhere non-negative,
since the distribution $f_0$ is;%
\footnote{
  Though this is a sensible physical requirement, it is not evident
  that it makes any qualitative difference in simulations of the
  hard-loop effective theory.
}
and (v) preferably $\Omega(\theta)$ should be monotonic for
$0 \le \theta \le \pi/2$.  The last condition is simply superstition:
We did not want to have to worry whether a multiple-hump
distribution might perhaps have qualitatively different behavior than
a single-hump one, and so we chose to study the simplest, most natural
case.  After some experimentation with distributions containing only
$l\le 6$ harmonics, we settled on
\begin {subequations}
\label {eq:fchoice}
\begin {equation}
   \Omega(\theta) \propto (\cos^2\beta - \cos^2\theta)^3 + \sin^6\beta
   \qquad
   \mbox{with $\beta=0.480$.}
\end {equation}
This corresponds to
\begin {equation}
  \Omega_{00} = 1 \,,
  \qquad
  \Omega_{20} = -0.790 \,,
  \qquad
  \Omega_{40} = 0.367 \,,
  \qquad
  \Omega_{60} = -0.093 \,,
\end {equation}
\end {subequations}
with all other $\Omega_{lm}=0$.
Fig.\ \ref{fig:fchoice} shows a plot of the angular dependence.

The dominant instability of this distribution is a Weibel instability
with wave-number $\k$ along the $z$ axis.  A perturbative analysis of
this instability (see Sec.\ \ref{sec:pert}) shows that the mode with
the maximum growth rate $\gamma$,
in the infinite-volume, continuum limit,
has
\begin {equation}
   k = 0.827 \, m_\infty ,
   \qquad
   \gamma = 0.273 \, m_\infty \,.
\label {eq:gamma}
\end {equation}


\subsection {Initial conditions}
\label {sec:initial}

We want to choose initial conditions that (i) have a random mix of
perturbatively unstable modes, and (ii) are insensitive to the
ultraviolet cutoff so that, for instance, the energy density has a good
continuum limit.  A simple choice that works in any dimension is
to choose Gaussian noise for the gauge fields $\A$ with an exponential
fall-off in $\k$:
\begin {equation}
   \A(\x) = \int \frac{d^d k}{(2\pi)^d} \, \A(\k) \, e^{i\k\cdot\x} ,
\end {equation}
where $d{=}1$ or 3 is the number of spatial dimensions and
the $A_i^a(\k)$ are Gaussian random variables with variance%
\footnote{
  On the lattice, we replace $k^2$ by $\sum_i \frac 4 {a^2}\sin^2(k_i a/2)$
  in (\ref{eq:initialA}) and define the initialization of link matrices by
  $U_i(\x) = \exp[i g a A_i^a(\x) \sigma^a/2]$.
}
\begin {equation}
   \langle A_i^a(\k)^* \, A_j^b(\k') \rangle
   = \left( \frac{\Delta}{g} \, e^{-k^2/k_0^2} \right)^2 \,
     \delta^{ab} \delta_{ij} \,
     (2\pi)^d \, \delta^{(d)}(\k-\k') .
\label {eq:initialA}
\end {equation}
Here $\Delta$ and $k_0$ are constants.
It is conventional in simulations of SU(2) gauge theory to rescale
the definition of fields to absorb factors of $2/g$, but in this paper we
will show all factors explicitly.%
\footnote{
  That is, our normalization
  convention will be the traditional, perturbative convention
  that $D_\mu = \partial_\mu - i g A_\mu^a \sigma^a/2$ in the
  fundamental representation, where the
  $\sigma^a$ are the Pauli matrices.
  The redefinition $A^\mu \to 2 A^\mu/g$ would instead make
  $D_\mu = \partial_\mu - i A_\mu^a \sigma^a$.
}
Our simulations are carried out in $A_0{=}0$ gauge, but the evolution
equations and all the observables we report are gauge invariant.
We choose
\begin {eqnarray}
  \E &=& -\dot\A = 0
\label {eq:initialE}
\\
\noalign{\hbox{and}}
  W &=& 0
\end {eqnarray}
as our remaining initial conditions.
Both are motivated solely by simplicity.  In particular,
(\ref{eq:initialE}) automatically implements Gauss' Law.

For small $\Delta$ (so that perturbation theory applies) in
three dimensions, these initial
conditions correspond to to an initial magnetic energy density of
\begin {equation}
   \langle \half B^2 \rangle
   \simeq \frac{3 \nu \, k_0^5}{64 \, (2\pi)^{3/2}}
   \, \frac{\Delta^2}{g^2}
\end {equation}
in the infinite-volume, continuum limit,
where $\nu = 6$ is the number of gauge boson degrees of freedom
[2 spin times 3 color for SU(2)].
In 1D+3V dimensions, the corresponding result is
\begin {equation}
   \langle \half B^2 \rangle
   \simeq \frac{\nu \, k_0^3}{16 \, (2\pi)^{1/2}}
   \, \frac{\Delta^2}{g^2} \,.
\end {equation}

In the 3D+3V dimensional simulations reported in this paper, we choose the
momentum cut-off scale $k_0$ in (\ref{eq:initialA}) as%
\begin {equation}
   k_0 = 2 m_\infty .
\end {equation}
Unless otherwise stated, we take the initial amplitude $\Delta$ 
in 3D+3V simulations to be
\begin {equation}
   \Delta = 0.02 \, m_\infty^{-1/2} .
\end {equation}
In 1D+3V simulations, we take $\Delta = 0.014 \, m_\infty^{+1/2}$,
which corresponds
to roughly the same value of the dimensionless%
\footnote{
  In $d$ spatial dimensions, the coupling $g$ has mass dimension
  $(3-d)/2$ and $g B$ has mass dimension 2.  Since $g$ can be scaled out
  of the classical equations of motion by $A \to A/g$, the natural
  dimensionless measure of energy for classical simulations is
  $g^2 \langle \half B^2 \rangle / m_\infty^4$ rather than, for
  instance, $\langle \half B^2 \rangle / m_\infty^{d+1}$.
}
energy
$g^2 \langle \half B^2 \rangle / m_\infty^4$.

To perform abelian simulations, we use our SU(2) simulation code
with initial conditions that lie in a single direction in adjoint
color space, {\it i.e.}
\begin {equation}
   \langle A_i^a(\k) \, A_j^b(\k') \rangle
   = \left( \frac{\Delta}{g} \, e^{-k^2/k_0^2} \right)^2 \,
     \delta^{a3} \delta^{b3} \delta_{ij} \,
     (2\pi)^d \, \delta^{(d)}(\k-\k') .
\end {equation}


\section {Additional Results}
\label{sec:more}

\begin{figure}[ht]
\includegraphics[scale=0.80]{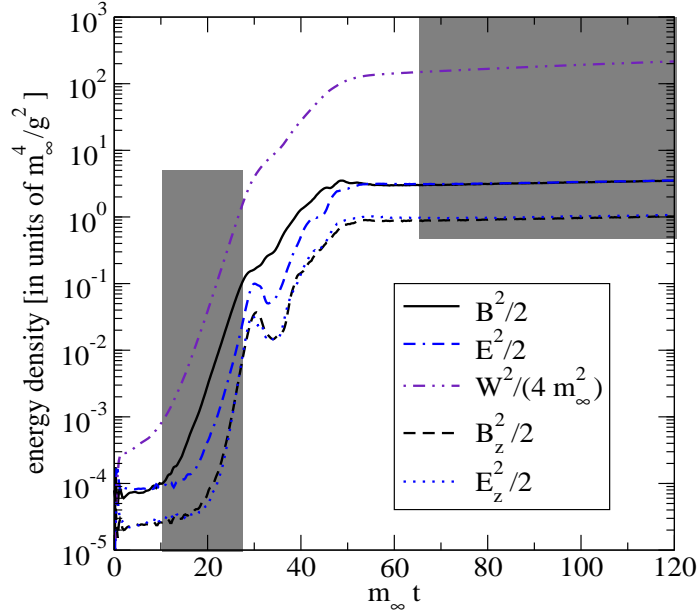}
\caption{%
    \label{fig:energies}
    Various components of the energy density for the same three dimensional
    non-abelian simulation as in Fig.\ \ref{fig:example}.
}
\end{figure}

\begin{figure}[ht]
\includegraphics[scale=0.45]{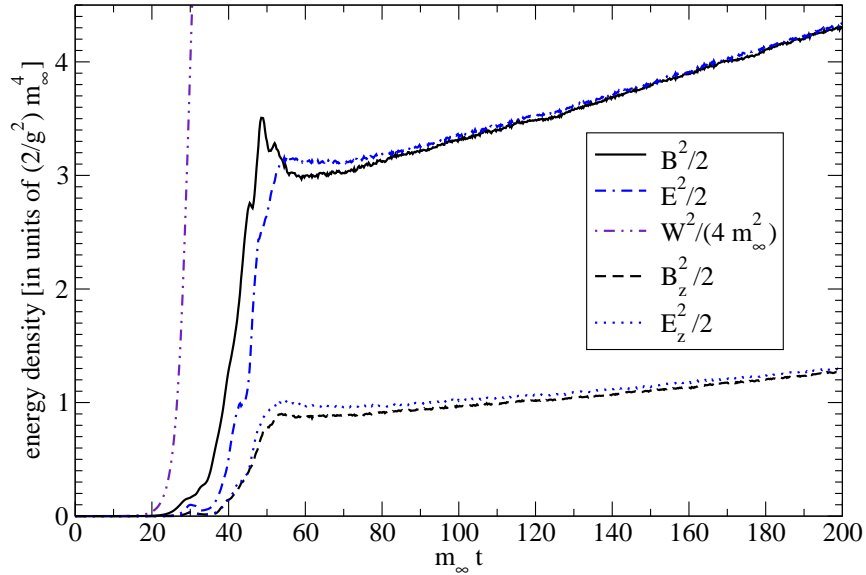}
\caption{%
    \label{fig:energies_linear}
    Same as Fig.\ \ref{fig:energies}, but with a linear vertical axis.
}
\end{figure}

In addition to the magnetic energy plotted in Figs.\ \ref{fig:example}
and \ref{fig:ex_linear}, it is interesting to see a breakdown of
other components of the energy of the soft fields.  For an
isotropic distribution $f_0$, the combination
\begin {equation}
  \int_\x \left[
       \half E^2 + \half B^2 + \fourth \, m_\infty^{-2} \int_\v W^2
  \right]
\label {eq:EBWenergy}
\end {equation}
would be conserved \cite{Wformalism}.
We will therefore loosely refer to $\fourth \, m_\infty^{-2} \int_\v W^2$
as the ``$W$ field energy density.''
Figs.\ \ref{fig:energies} and \ref{fig:energies_linear}
compare various components of the
volume-averaged energy density
as a function of time, including the magnetic, electric, and
$W$ field energy densities.
As one indication of the anisotropy of the
soft fields, we also show $\half B_z^2$ and $\half E_z^2$.
Note that, unlike Fig.\ \ref{fig:example}, we now show time all the
way back to the initial conditions at $t=0$.
There is a very early transient at $m_\infty t \lesssim 1$ that
is difficult to see in the plot, when the initial energy in the magnetic
fields is quickly shared with the other degrees of freedom, $E$ and $W$.
The unstable modes start to grow, but it is only when they grow
large enough to dominate the energy density that this manifests
as the start of exponential growth in the magnetic energy, around
$m_\infty t \sim$ 10 to 15.

\begin{figure}[ht]
\includegraphics[scale=0.80]{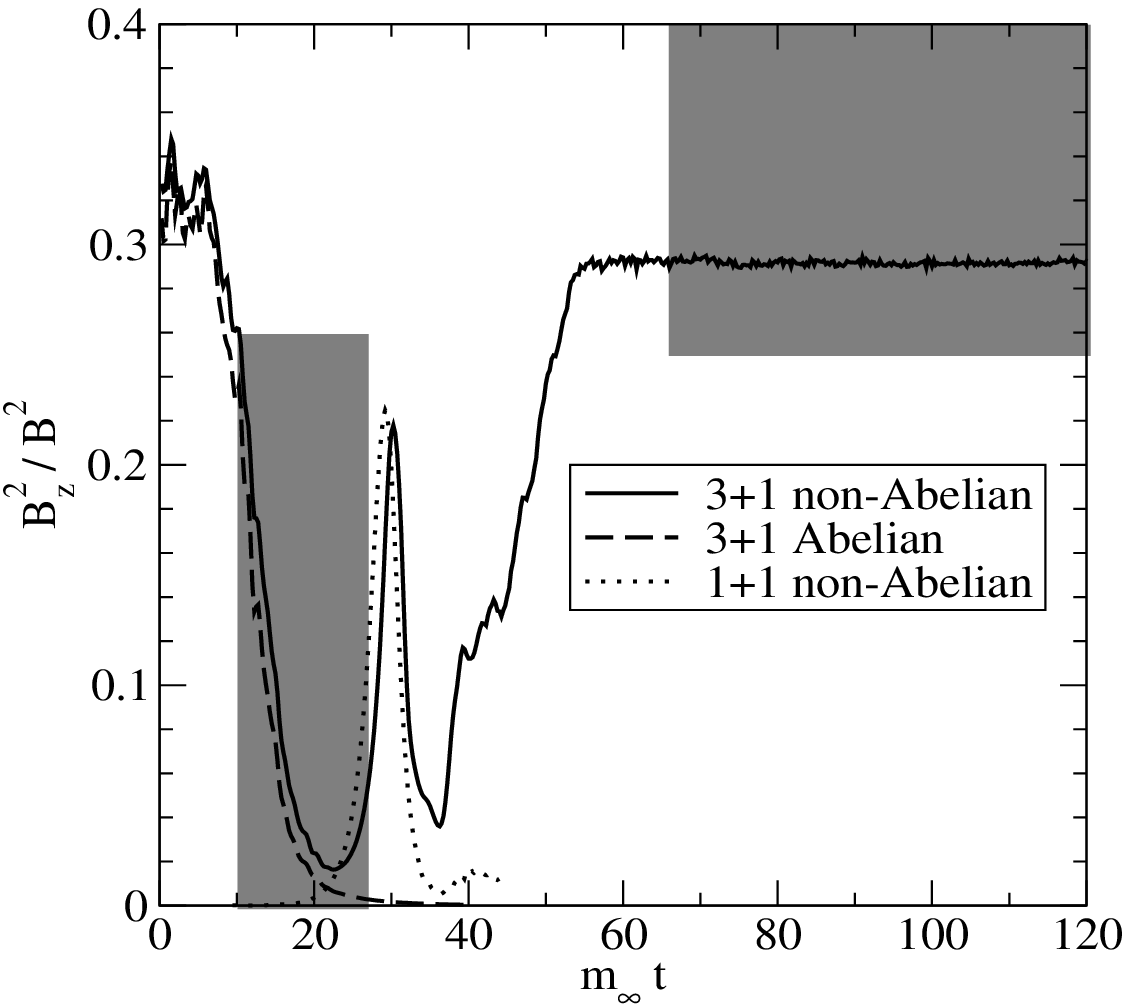}
\caption{%
    \label{fig:Bzratio}
    The ratio of the $B_z^2$ contribution to the magnetic energy
    density to the total $B^2$ for the 3+1 and 1+1 dimensional
    simulations of Fig.\ \ref{fig:example}.  To make the curves line
    up, the time
    origin of the 1+1 dimensional simulation has been shifted just as
    in  Fig.\ \ref{fig:example}
    ($m_\infty t \to m_\infty t + 9.2$).
}
\end{figure}

Fig.\ \ref{fig:Bzratio} shows the ratio of the $B_z^2$
contribution to the magnetic energy density to the total $B^2$.
If the soft fields were isotopic, this ratio would be $1/3$.
This is indeed the value at the earliest times, due to our isotropic
soft-field initial conditions.  For $10 \lesssim m_\infty t \lesssim 20$,
as the Weibel instability first starts to grow and dominate the
magnetic field, $B_z$ drops dramatically
compared to $B$.  This is because the Weibel instability is dominated by
modes with wavenumber $\k$ along the $z$ axis, and magnetic fields are
perpendicular to $\k$.  In the abelian case, that is the end of the
story: $B_z$ continues to become more and more insignificant as the
unstable modes grow.  In the non-abelian case,
$B_z = (\grad \times \A)_z - ig [A_x, A_y]$, and the non-abelian
commutator contributes even when $\k$ is in the $z$
direction.  This contribution to $B_z$ will grow with time, as
the gauge fields $A_x$ and $A_y$ grow due to the instability,
and $B_z$ should become the same order of magnitude as $B_x$ and
$B_y$ once those fields become non-perturbatively large
(\ref{eq:Bna}).  This growth corresponds to roughly
$23 < m_\infty t < 30$ in the figure.  At $m_\infty t \sim 30$, the ratio
is no longer small, and so one would expect this to be where some
deviation from the behavior of abelian instabilities should occur.
This was indeed the case in the non-abelian
magnetic energy curve of Figs.\
\ref{fig:example} and \ref{fig:energies}, where one sees a first small
bump in the plot of magnetic energy vs.\ time at $m_\infty t \sim 30$.

Based on arguments concerning 1+1 dimensional configurations of fields,
it was conjectured in Ref.\ \cite{AL} that dynamics at the non-abelian
scale would cause the field configurations to approximately abelianize,
and that the Weibel instability would then again take over, causing
the field to grow all the way to the ultimate scale (\ref{eq:Ba}).
If abelian-like Weibel modes come to dominate, then $B_z$ should
again become small relative to $B$.  This was seen in numerical
simulations by Rebhan {\it et al}.\ and is reproduced in our own
1+1 dimensional data in Fig.\ \ref{fig:Bzratio} for $m_\infty t > 30$.
For a brief time, the $3+1$ dimensional simulations behave similarly,
but the decrease of $B_z/B$ eventually stops and reverses at
$m_\infty t \sim 36$.  The ratio then starts growing again and
finally levels out near 0.29 (slightly lower than one third) at
$m_\infty t \sim 53$, which is near the beginning of the linear growth of
magnetic energy in Figs.\ \ref{fig:ex_linear} and
\ref{fig:energies_linear}.

\begin{figure}[ht]
\includegraphics[scale=0.80]{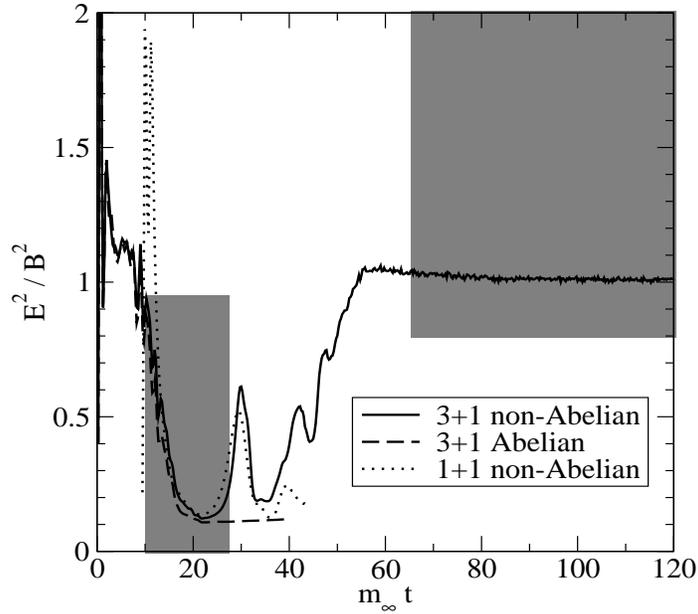}
\caption{%
    \label{fig:EBratio}
    The ratio of electric to magnetic energy.
}
\end{figure}

In Fig.\ \ref{fig:EBratio}, we plot the ratio of electric energy to
magnetic energy.  The behavior is somewhat similar to the previous plot.
Looking at the abelian curve (dashed), we see that, as the Weibel instability
grows and dominates the energy density, the ratio drops and eventually
asymptotes to a constant.  Theoretically, this constant should be
$(\gamma/k)^2$, where $\gamma$ and $k$ are the growth rate and wave number
of the dominant unstable mode.  Using (\ref{eq:gamma}), the
infinite-volume, continuum value is $(\gamma/k)^2 = 0.11$,
which is in good agreement with the figure.  For the non-abelian
simulations, however, the ratio subsequently
grows as non-abelian interactions first
become important, and then decreases again,
consistent with the picture of approximate abelianization.
For 3+1 dimensional
non-abelian simulations, however, the ratio
then increases yet again, and finally approaches
unity in the linear energy growth regime.  Note that
plasma oscillations with momenta large compared to the mass
$m_\infty$ would naturally have $E^2{\simeq}B^2$.  Fig.\ \ref{fig:EBratio}
might therefore be a hint about what sort of soft field
excitations dominate the energy in the linear regime.

\begin{figure}
\includegraphics[scale=0.80]{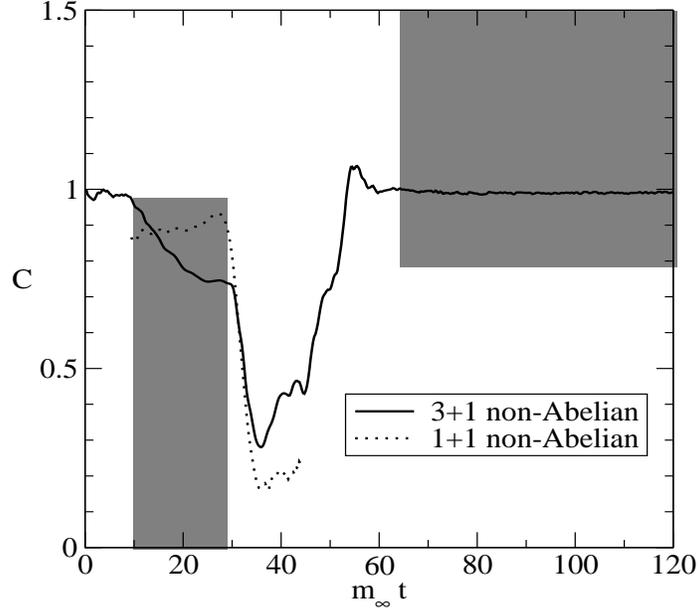}
\caption{%
    \label{fig:Jcomm}
    The local measure (\ref{eq:C}) of the relative size of commutators,
    plotted as a function of time,
    for the non-abelian simulations of Fig.\ \ref{fig:example}.
}
\end{figure}

It is interesting to directly address
the conjecture of Ref.\ \cite{AL}
that non-perturbative dynamics would
abelianize the gauge field configurations.
We consider the following local observable, related to ones used
by Refs.\ \cite{AL,RRS}:
\begin {equation}
  C \equiv
  \frac3{\sqrt2} \>
  \frac{
    \left[
      \int \frac{d^3x}{V} \>
      \Bigl( ([j_x,j_y])^2 + ([j_y,j_z])^2 + ([j_z,j_x])^2 \Bigr)
    \right]^{1/2}
  }{
    \int \frac{d^3x}{V} \>
    |\j|^2
  }
  \,,
\label {eq:C}
\end {equation}
where
$[j_x,j_y]^2 \equiv \epsilon^{abc} j^b_x j^c_y \, \epsilon^{amn} j^m_x j^n_y$,
{\em etc}.
The normalization of $C$ has been chosen so that
$C$ would be unity if the components of $\j$ were independent random
numbers with the same distribution.  For an abelian configuration,
$C$ would be zero.  Fig.\ \ref{fig:Jcomm} shows the time development
of $C$ in our canonical non-abelian simulations.  We see that $C$ drops
suddenly when non-abelian interactions first become important, in
agreement with the abelianization conjecture.  However, in the
3+1 dimensional simulations, $C$ later rises again all the way to
unity, showing no local abelianization in the linear growth regime.

In Figs.\ \ref{fig:Bzratio}--\ref{fig:Jcomm}, we have not
displayed any data for the 1+1 non-abelian or 3+1
abelian simulations at late times when the corresponding energy curves
in Fig.\ \ref{fig:example} grew so large that they are
approaching the top of the plot.
At those and later times, the fields in our simulations
grow so large that they
become sensitive to the discretization of the lattice, and the behavior
of the results is then a lattice artifact.
We shall discuss this issue more thoroughly in Sec.\ \ref{sec:a}.

\begin{figure}[ht]
\includegraphics[scale=0.80]{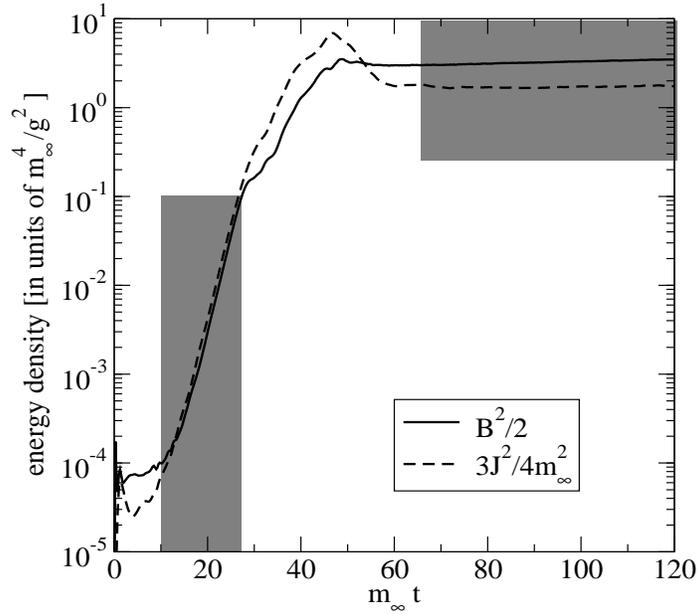}
\caption{%
    \label{fig:W1sqr}
    A comparison of magnetic energy density (solid)
    and the average squared current $|\j|^2$ (dashed), with the
    latter normalized as in (\ref{eq:W1energy}),
    for the 3+1 dimensional non-Abelian
    simulation of Fig.\ \ref{fig:example}.
}
\end {figure}

\begin{figure}[ht]
\includegraphics[scale=0.80]{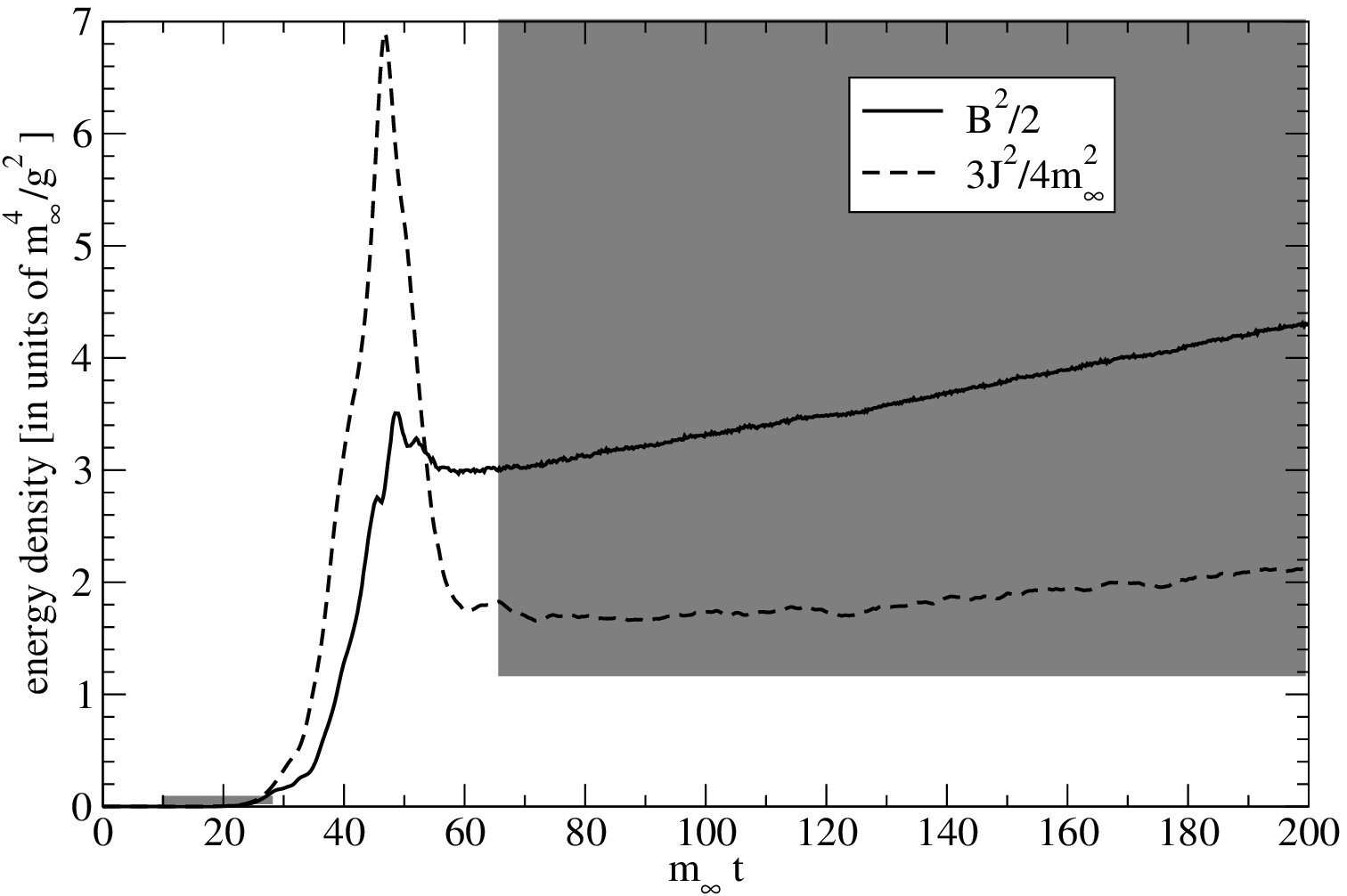}
\caption{%
    \label{fig:W1sqr_linear}
    Same as Fig.\ \ref{fig:W1sqr}, but with a linear vertical axis.
}
\end{figure}

In 1+1 dimensional simulations, the size of particle
currents $\j$ was used by Rebhan {\it et al.}\ \cite{RRS}
to track the growth of the instability.
In Figs.\ \ref{fig:W1sqr} and \ref{fig:W1sqr_linear},
we track a related quantity for our 3+1 dimensional simulations.
These figures
compare the magnetic energy density to the energy
density in the $l{=}1$ components of $W$.  The latter is given by
(\ref{eq:EBWenergy}) as $W_{1m}^a W_{1m}^a/(4 m^2_\infty)$, which is
directly proportional to the squared current $|\j|^2$ by (\ref{eq:j}):
\begin {equation}
   \frac{1}{4 m_\infty^2} \sum_m W_{1m}^a W_{1m}^a
   = \frac{3 |\j|^2}{4 m_\infty^2} \,.
\label {eq:W1energy}
\end {equation}
Fig.\ \ref{fig:W1sqr_linear} shows that
$\frac 34 \, |\j|^2/m_\infty^2$ exceeds the magnetic energy
$\half \B^2$ by roughly a factor of two in the exponential growth regime,
but then drops substantially and is only about half the magnetic energy
in the linear regime.


\section {Checks and systematic errors}
\label {sec:checks}


\subsection {Perturbative growth rate}
\label {sec:pert}

One simple check of simulations is to see whether the rate of
exponential energy growth in the perturbative regime
is consistent with the infinite-volume,
continuum prediction for the Weibel instability growth rate.
To compute the perturbative growth rate, one looks for exponentially
growing solutions ($\Im\omega > 0$)
to the linear dispersion relation
\begin {equation}
   [(-\omega^2 + k^2) g^{\mu\nu} - K^\mu K^\nu + \Pi^{\mu\nu}(\omega,\k)]
   A_\nu = 0 ,
\end {equation}
where $K = (\omega,\k)$ and
$\Pi$ is the hard loop self-energy \cite{mrow0,randrup&mrow}
\begin {eqnarray}
   \Pi^{\mu\nu}(\omega,\k) &=&
   g^2 t_R \int_\p
     \frac{\partial f_0(\p)}{\partial p^{\,l}}
     \left[ -v^\mu g^{l\nu}
            + \frac{v^\mu v^\nu k^l}{-\omega+\v\cdot\k-i\epsilon}
     \right] .
\label {eq:Pi}
\end {eqnarray}
In Appendix \ref{app:pi}, we formulate this in terms of the angular
distribution
${\Omega(\v)}$ and then find a result that is well suited for
calculations of growth rates in cases where $\Omega$ is expressed
in terms of spherical harmonics.
Specifically, in situations where the hard particle distribution
$f_0(\p)$ is axi-symmetric, the transverse self-energy for
wave-vectors along the $z$ axis can be expressed as
\begin {subequations}
\label{eq:Piperp}
\begin {equation}
  \Pi_\perp(\omega, k {\bm e}_z)
  =
  \half m_\infty^2 \sum_l \sqrt{2l{+}1}\;
  \kappa_l\!\left(\frac{\omega}{k}\right) \, {\Omega_{l0}}
\end {equation}
with
\begin {equation}
  \kappa_l(\eta) \equiv
  (1+\eta^2) \delta_{l0}
  + (1-\eta^2) [(l+1) Q_{l+1}(\eta) - (l-1) \eta Q_l(\eta)] .
\end {equation}
\end {subequations}
Here, $Q_l(\eta)$ is the Legendre function of the second kind defined
so that it is regular at $\eta=\infty$ and the cut is chosen to
run from $-1$ to +1.%
\footnote{
  For example,
  $
    Q_0(z) = \frac12 \ln \frac{z+1}{z-1}
  $,
  $
    Q_1(z) = \frac{z}2 \ln \frac{z+1}{z-1} - 1
  $,
  and
  $
    Q_2(z) = \frac{(3z^2-1)}4 \> \ln \frac{z+1}{z-1} - \frac{3z}{2}
  $.
}
In cases where the dominant instability is a Weibel instability with
$\k$ along the axis of symmetry, the dispersion relation then reduces
to
\begin {equation}
   -\omega^2 + k^2 + \Pi_\perp(\omega,k {\bm e}_z) = 0 .
\end {equation}
For a given distribution $\Omega(\theta)$, one can solve this equation
numerically for each $k$, and then scan over $k$ to look for the
mode with the largest growth rate $\gamma = \Im\omega$.
For the distribution
(\ref{eq:fchoice}) used in our simulations, the resulting
$\gamma$ and $k$ were given earlier in Eq.~(\ref{eq:gamma}).

\begin{figure}[ht]
\includegraphics[scale=0.45]{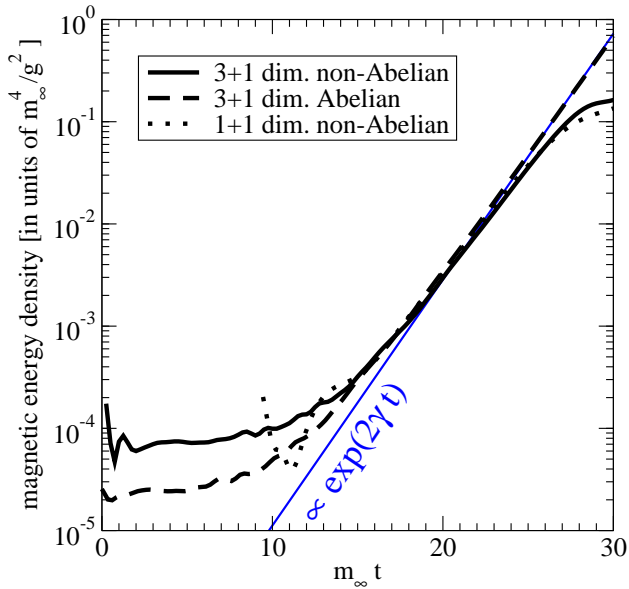}
\caption{%
    \label{fig:gamma}
    The perturbative growth in Fig.\ \ref{fig:example} compared to
    the predicted growth (\ref{eq:Bgamma}) of the dominant Weibel mode
    (\ref{eq:gamma}).  As in previous figures, we have shifted the time
    origin of the 1+1 dimensional simulation to line
    up the curves ($m_\infty \to m_\infty t + 9.2$).
}
\end{figure}

The rate $\gamma$ gives the growth rate of the perturbative vector potentials.
The corresponding magnetic energy should grow as the square of
the field strength, so that
\begin {equation}
   \half B^2 \propto e^{2\gamma t} .
\label {eq:Bgamma}
\end {equation}
In Fig.\ \ref{fig:gamma}, we again show our canonical simulations
of Fig.\ \ref{fig:example}, this time focusing on the perturbative
regime and comparing the slope of $\ln(\half B^2)$ to $2\gamma$.
We get reasonable agreement.  Keep in mind that
there is an entire spectrum of unstable
modes, not just the dominant mode discussed above;
so the exponential growth of instabilities is not described by
a single exponential at early times.


\subsection {Finite volume errors}

\begin{figure}[ht]
\includegraphics[scale=0.45]{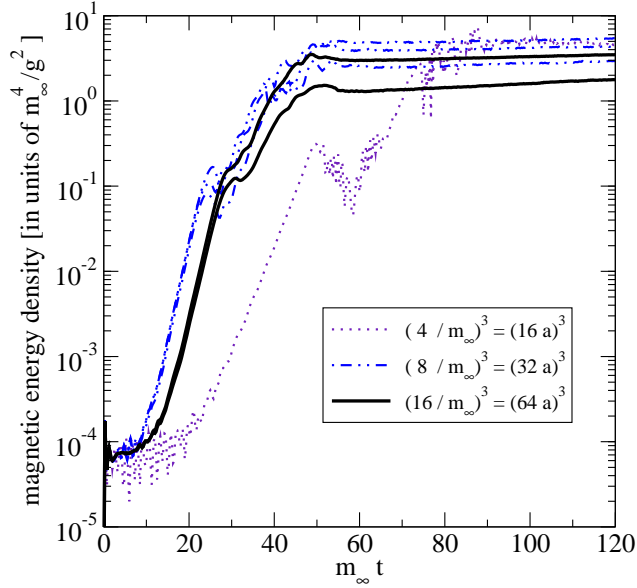}
\caption{%
    \label{fig:volume}
    Magnetic energy vs.\ time as in Fig.\ \ref{fig:example}, but
    showing results for several different physical volumes.
    Multiple lines for a single volume correspond to different
    instantiations of the random initial conditions.
}
\end{figure}

\begin{figure}[ht]
\includegraphics[scale=0.45]{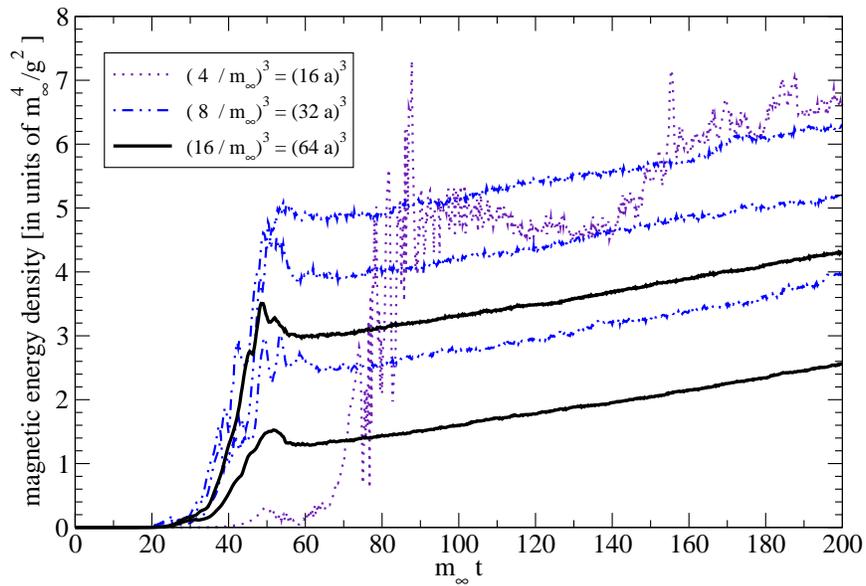}
\caption{%
    \label{fig:volume_linear}
    Same as Fig.\ \ref{fig:volume}, but with a linear vertical axis.
}
\end{figure}

In Figs.\ \ref{fig:volume} and \ref{fig:volume_linear},
we show how our 3+1 dimensional non-abelian simulation
results of Fig.\ \ref{fig:example} change if we decrease the physical
volume.
When comparing results for different volumes,
it is important to realize that different
simulations will have different random initial conditions---there is
no good way to start two simulations with the same initial
conditions when they have different physical volumes.
To get an idea of the size of this effect, we have simulated various
volumes more than once, changing the seed of our random number
generator.  For very large volumes, one would expect the spread of
the results to decrease with increasing volume because the volume
average taken in computing the energy density averages over multiple
spatial regions with different initial conditions.

As one can see, the simulations shown in Fig.\ \ref{fig:volume_linear}
still show a significant variation in results
with respect to initial conditions at
a volume of $L^3 = (16/m_\infty)^3$.  However, the {\it rates}\/ of
growth in the perturbative, exponential regime and the non-perturbative,
linear regime have clearly reached their large volume limits.
Specifically, Fig.\ \ref{fig:volume} shows that the perturbative growth
rate (the slope of the curve for $10 \lesssim m_\infty \lesssim 20$)
is not significantly affected in increasing volume from
$(8/m_\infty)^3$ to $(16/m_\infty)^3$.
And Fig.\ \ref{fig:volume_linear} similarly shows little effect on
the slope of the late-time linear growth behavior.  This implies that
the phenomenom of linear growth is not a finite-volume artifact.



\subsection {Finite lattice spacing errors}
\label{sec:a}

\subsubsection {3+1 dimensions}

\begin{figure}[ht]
\includegraphics[scale=0.45]{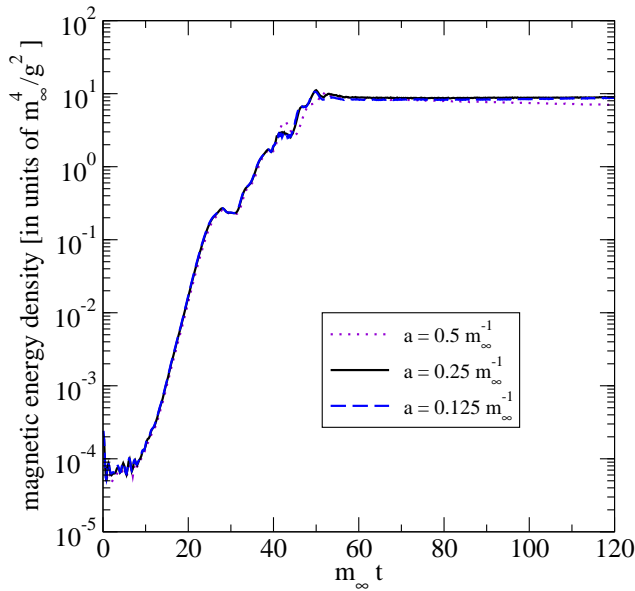}
\vspace*{-10pt}
\caption{%
    \label{fig:spacing}
    Magnetic energy vs.\ time as in Fig.\ \ref{fig:example}, but
    showing results for several different choices of lattice spacing
    for a volume $L^3 = (8/m_\infty)^3$.
    The $a{=}0.25 \, m_\infty^{-1}$ and $a{=}0.125\, m_\infty^{-1}$
    curves are difficult to distinguish on this plot.
}
\end{figure}

Fig.\ \ref{fig:spacing} shows the lattice spacing dependence of our
simulations for a lattice volume of $L^3 = (8/m_\infty)^3$.
We chose a smaller volume than that used in our canonical simulation
of Fig.\ \ref{fig:example} so that we could push to smaller
lattice spacing with our available computer resources.  Recall from
Fig.\ \ref{fig:volume} that this smaller volume is
adequate for reproducing the linear growth rate.
Here and in all our simulations, the time step used in evolving the
system is $\delta t = 0.1 \, a$.\thinspace%
\footnote{
   Reducing $a$ therefore also reduces the time step.  In addition,
   for one of our simulations,
   we checked that holding $a$ fixed
   and reducing $\delta t$ to $0.05 \, a$ makes little ($< 5$\%)
   difference to the result.  In particular, the continued growth of energy in
   the linear regime does not appear to be a time discretization artifact.
} 

\begin{figure}[ht]
\includegraphics[scale=0.45]{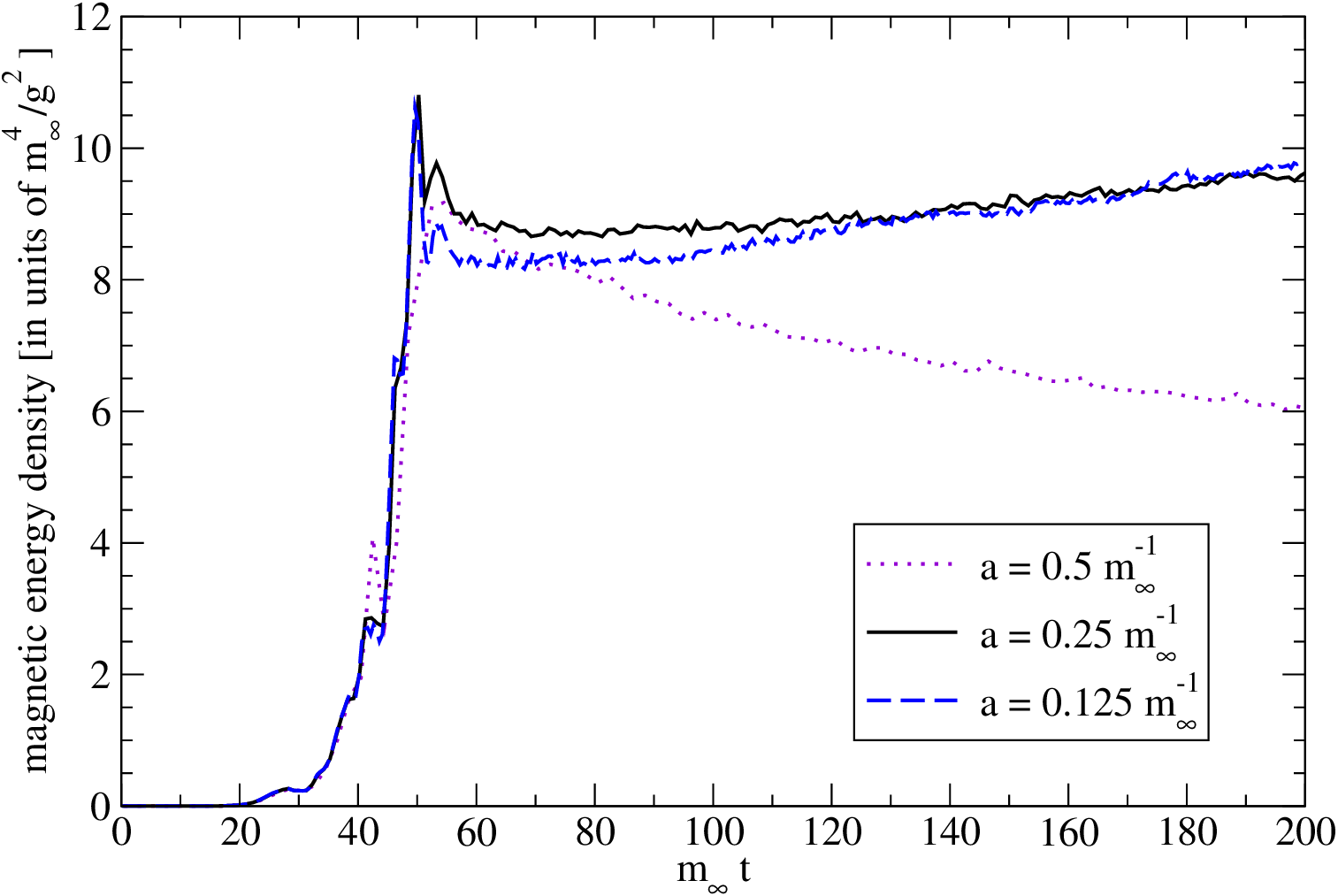}
\vspace*{-10pt}
\caption{%
    \label{fig:spacing_linear}
    Same as Fig.\ \ref{fig:spacing}, but with a linear vertical axis.
}
\end{figure}

Since we have already seen that different initial conditions can
produce substantially different behaviors at this volume, especially
at the transition between exponential and linear growth, it is
essential in comparing lattice spacings to ensure that the initial
conditions are as nearly identical as possible.  In Figures
\ref{fig:spacing} and \ref{fig:spacing_linear}, we have done this by
drawing the smallest lattice spacing configuration randomly as
described in subsection \ref{sec:initial}, and converting it into
larger lattice spacing configurations via blocking.  Besides the
lattice spacing and volume, we otherwise use the same ``canonical''
values for variables as in most previous simulations.

From Fig.\ \ref{fig:spacing}, one sees negligible spacing dependence
in the early, perturbative growth of the instability.
From Fig.\ \ref{fig:spacing_linear}, we conclude that spacing does
not have a significant effect effect on the {\it slope}\/ of the linear
growth in magnetic energy provided $a$ is at least as small as our
canonical value of $0.25 \, m_\infty^{-1}$.  The largest lattice
spacing shows approximately the same exponential growth behavior, but
different linear growth behavior.  This is apparently because there is
more energy in ultraviolet degrees of freedom during the linear growth
period, and these degrees of freedom are compromised by the large
lattice spacing.

\begin{figure}[ht]
\includegraphics[scale=0.40]{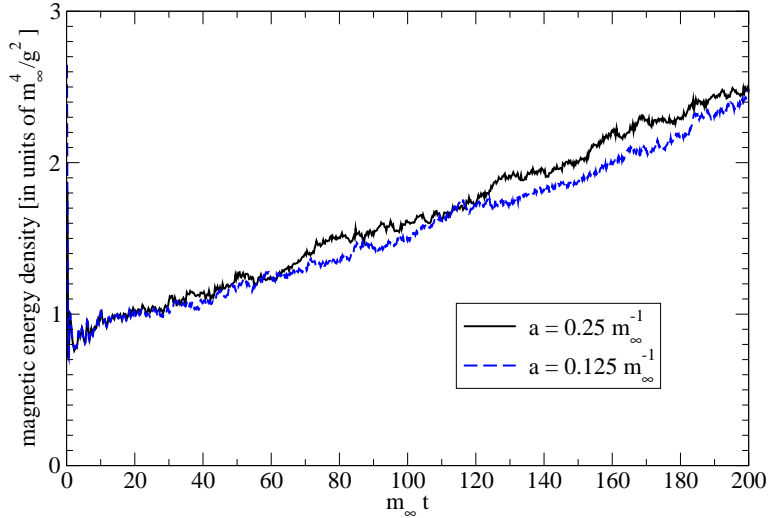}
\vspace*{-10pt}
\caption{%
    \label{fig:spacingBigDelta_linear}
    Lattice spacing dependence for non-perturbatively large initial
    conditions, $\Delta = 2$.
}
\end{figure}

To test dependence on both lattice spacing and initial amplitude $\Delta$, 
we show in Fig.\ \ref{fig:spacingBigDelta_linear} what happens if we
start with large, non-perturbative initial conditions%
\footnote{
  \label {foot:Ak}%
  For historical reasons concerning the development of our methods, this
  data was produced with a slightly different procedure
  for matching initial conditions between different lattice spacings:
  initial conditions for coarser lattices were obtained
  from those for finer lattices by setting Fourier amplitudes
  $\A_\k$ exactly the same
  for the physical momenta $\k$ that exist on the coarser lattice
  (choosing $k$ so that $-\pi/a < k < \pi/a$).
}
of
$\Delta = 2.0$.  Here the system goes directly into linear
growth, and the $a=0.25 \, m_\infty^{-1}$
and $a=0.125 \, m_\infty^{-1}$ results are reasonably close.

From the consistency of the slope of the linear regime for
$a \le 0.25 \, m_\infty$, we conclude that the existence of the
linear growth regime is not an artifact of finite lattice-spacing.


\subsubsection {1+1 dimensions}

\begin{figure}[ht]
\includegraphics[scale=0.45]{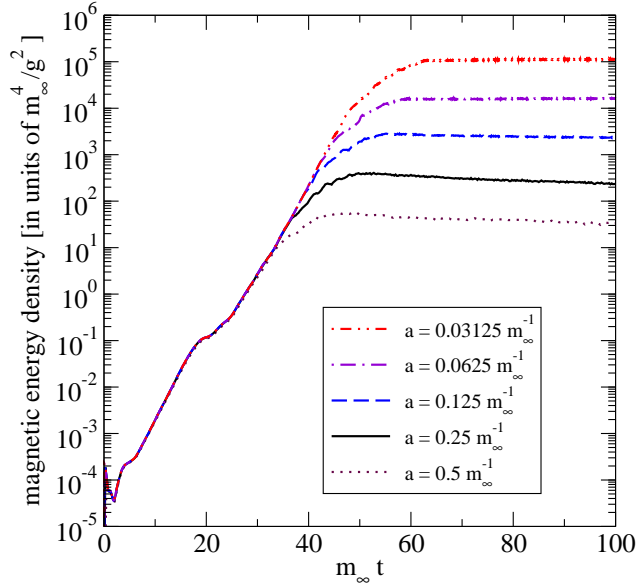}
\vspace*{-5pt}
\caption{%
    \label{fig:1Dspacing}
    Magnetic energy vs.\ time for 1+1 dimensional simulations with
    several different choices of lattice spacing
    for a system length $L = 128/m_\infty$.
    The remaining parameters are as in Fig.\ \ref{fig:example}.
    except that here we have not shifted the origin of the time axis.
}
\end{figure}

To sharpen our conclusions that the linear growth regime in
3+1 dimensions is not
an artifact of our simulations, it is useful to look at a different
case, where the end of exponential growth {\it is}\/ a lattice
artifact.
Such an example is provided by our 1+1 dimensional simulations,
shown in Fig.\ \ref{fig:1Dspacing} for a variety of lattice spacings.%
\footnote{
  Again for historical reasons,
  initial conditions were matched using the method of footnote
  \ref{foot:Ak}.
}
We display results to
much larger energies than in Fig.\ \ref{fig:example}.
Superficially, each individual curve looks vaguely similar to our
3+1 dimensional results: each shows an eventual end to exponential
growth, although at much higher energy density than in
3+1 dimensions.

In our simulations, we implement 1+1 dimensions by using our
3+1 dimensional code on a periodic lattice that is a single lattice spacing
wide in the $x$ and $y$ directions.  
No matter how fine the lattice is compared to the wavelength of the
unstable modes, the continuum limit will break down when the fields
become so large that the magnetic energy per plaquette is of order
its maximum possible value on the lattice, corresponding to
$\half B^2 \sim 1/(g^2 a^4)$.  Even at lower fields, the dynamics is
modified by ``irrelevant'' operators induced by the
lattice, such as $a^{2n} B^{n+1}$,
which become more and more important as $B$ grows larger.
As one takes the lattice spacing $a$ smaller and smaller,
the fields should be able to
grow larger and larger before these problems arise.
The termination of exponential growth in Fig.\ \ref{fig:1Dspacing}
clearly shows this behavior, demonstrating it is a lattice artifact.

\begin{figure}[ht]
\includegraphics[scale=0.45]{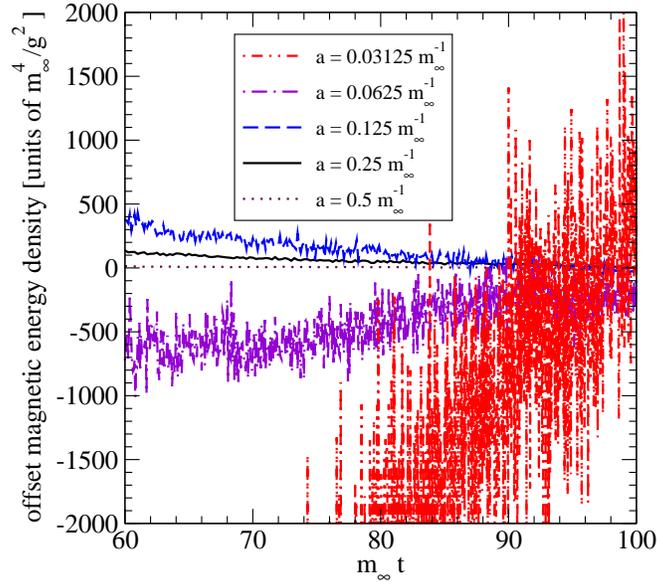}
\vspace*{-5pt}
\caption{%
    \label{fig:1Dspacing.slope}
    Same as Fig.\ \ref{fig:1Dspacing}, but the vertical axis is linear and
    different vertical offsets have been added to each curve to fit
    them onto the plot.  From top to bottom on the left side of the
    plot, the curves correspond to $a m_\infty =$ 0.125, 0.25, 0.5,
    0.0625, and 0.03125.
}
\end{figure}

In discussing our 3+1 dimensional simulations, we have emphasized how
the slope of the late-time linear growth is not significantly sensitive
to decreasing the lattice spacing.  In contrast, a similar look at the slope
of the phony late-time behavior of the 1+1 simulations, in
Fig.\ \ref{fig:1Dspacing.slope},
shows no such spacing independence.

\begin{figure}[ht]
\includegraphics[scale=0.45]{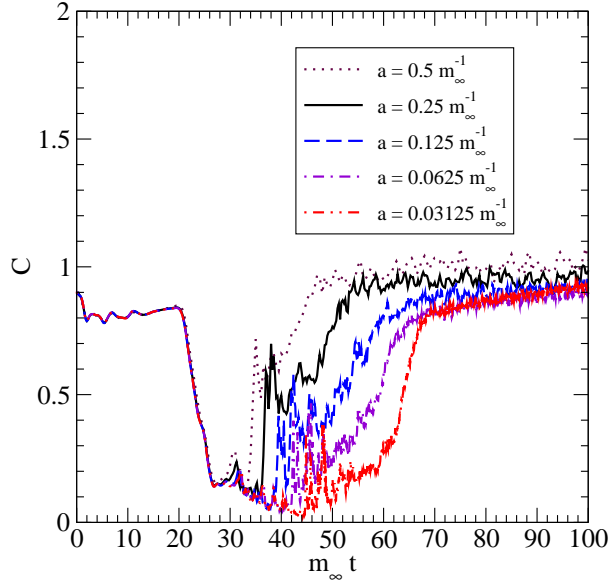}
\vspace*{-5pt}
\caption{%
    \label{fig:1DJcomm}
    The local measure (\ref{eq:C}) of the relative size of commutators
    for the various 1+1 dimensional simulations of Fig.\
    \ref{fig:1Dspacing}.  From top to bottom, the curves are in
    descending order in the size of $a$.
}
\end{figure}

Fig.\ \ref{fig:1DJcomm} shows the measure $C$
of local commutators, defined by Eq.~(\ref{eq:C}), for the 1+1
dimensional simulations of Fig.\ \ref{fig:1Dspacing}.  The
reader may now see late-time behavior that we did not display
in Fig.\ \ref{fig:Jcomm}, and also see that this behavior is a lattice
spacing artifact.  This is the reason we truncated our
1+1 dimensional curves in
Figs. \ref{fig:Bzratio}--\ref{fig:Jcomm}.
Specifically,
we truncated those curves when the magnetic energy densities reached
$20 \, m_\infty^4/g^2$, which is a little before where
the $a=0.25\, m_\infty^{-1}$ energy curve deviates from the
continuum limit in Fig.\ \ref{fig:1Dspacing} and where
the corresponding curve for $C$ suddenly begins to rise in
Fig.\ \ref{fig:Jcomm}.  A similar truncation was made for the Abelian
results, where a similar large-field issue arises since
we implement
compact rather than non-compact Abelian gauge theory.

\subsection {Finite \boldmath$\lmax$ errors}

\begin{figure}[t]
\includegraphics[scale=0.45]{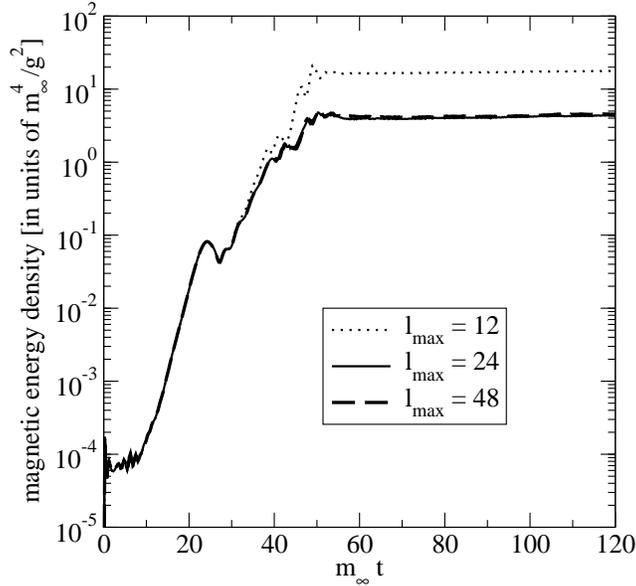}
\caption{%
    \label{fig:lmax32}
    Magnetic energy vs.\ time for several different choices of $\lmax$.
    These simulations have 
    $a = 0.25 \, m_\infty^{-1}$, $\lmax = 24$,
    $L^3 = (32 \, a)^3 = (8 \, m_\infty^{-1})^3$,
    and $\Delta=0.02\,m_\infty^{-1/2}$.
    The $\lmax{=}24$ and $\lmax{=}48$ curves are difficult to
    distinguish on this plot.
}
\end{figure}

\begin{figure}[ht]
\includegraphics[scale=0.40]{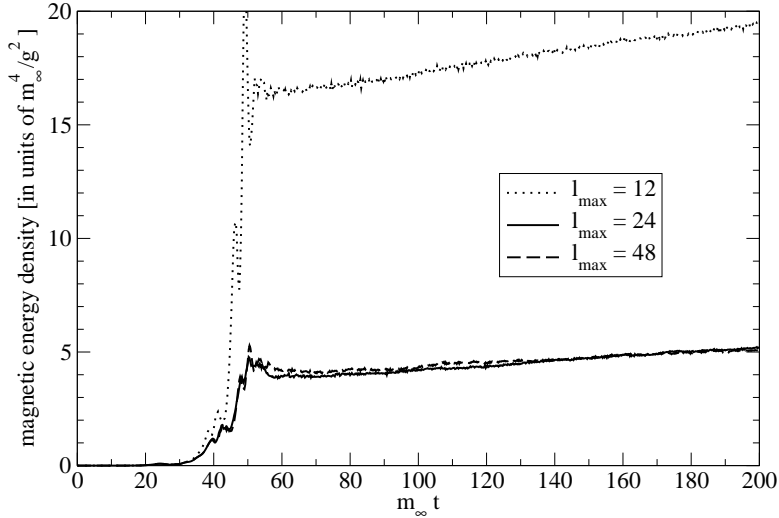}
\caption{%
    \label{fig:lmax32_linear}
    Same as Fig.\ \ref{fig:lmax32}, but with a linear vertical axis.
    The $\lmax{=}24$ and $\lmax{=}48$ curves are again very close together.
}
\end{figure}

Figs.\ \ref{fig:lmax32} and \ref{fig:lmax32_linear} show the dependence
of our results on $\lmax$.  The figures show simulations with fixed
lattice spacing $a = 0.25 \, m_\infty$ and fixed physical volume
$L^3 = (8/m_\infty)^3$.
By comparison of the $\lmax{=}24$ and
$\lmax{=}48$ curves in Fig.\ \ref{fig:lmax32_linear}, we conclude
that our canonical choice of $\lmax=24$ seems to be a good approximation
to the large $\lmax$ limit and that the linear growth regime is not
a finite $\lmax$ artifact.

We have chosen to run simulations only for even values of $\lmax$,
based on the
experience of Ref.\ \cite{BMR} for isotropic particle distributions,
where it was found that convergence to the large $\lmax$ limit was
much faster for $\lmax$ even than $\lmax$ odd.


\section {Discussion}
\label {sec:discussion}

We have found that 3D+3V dimensional non-abelian 
plasma instabilities behave
qualitatively differently than 1D+3V dimensional instabilities once they
grow non-perturbatively large.  Initially, there is a period of
non-perturbative growth that looks quite similar to the 1D+3V case
and to early conjectures about abelianization,
but eventually the 3D+3V instabilities settle into a period
characterized by linear rather than exponential
growth of the magnetic energy.
Based on our analysis of sources of systematic
error, we believe that this conclusion is not a simulation artifact.

There are many more things one would like to know about the linear growth
regime, such as its efficiency at scattering and isotropizing
typical particles in the plasma, its power spectrum, and possible
models for the underlying physical processes.
It would also be useful to check whether non-perturbative linear
growth occurs for
a wide variety of different distributions $f_0(\theta)$, beyond
the single case studied here.
We leave this and further
characterization to future work.


\begin{acknowledgments}

This work was supported, in part, by the U.S. Department
of Energy under Grant Nos.~DE-FG02-96ER40956 
and DE-FG02-97ER41027, by the National Sciences and Engineering
Research Council of Canada, and by le Fonds Nature et Technologies du
Qu\'ebec.

\end{acknowledgments}


\appendix

\section {Hard loop effective theory in \boldmath$lm$-space}
\label {app:M}

First, we will fill in a few steps in the text.  The transition from
(\ref{eq:basic2a}) to (\ref{eq:W1}) follows by using (\ref{eq:gradv})
to write
\begin {equation}
   \grad_\p f_0 =
   \frac{1}{p}\, \grad_\v f_0
   + \v \, \frac{\partial f_0}{\partial p} \,.
\end {equation}
Then integrate the last term by parts in $dp$ to obtain
\begin {equation}
    \int_0^\infty \frac{4\pi p^2 \>dp}{(2\pi)^3} \> \grad_{\p} f_0
    =
    (\grad_\v - 2\v) {\cal M}
\end {equation}
and so (\ref{eq:W1}).
Using (\ref{eq:gradv}) and (\ref{eq:L}), one may derive
that
\begin {equation}
   [\v,L^2] = 2(\grad_\v - \v) ,
\end {equation}
which yields (\ref{eq:W2}).

Now use the $lm$-expansions (\ref{eq:lmexpand}) of $W$ and $\Omega$
in (\ref{eq:W2}) and project out the $lm$-th component of the result.
Working in $A_0{=}0$ gauge, which is the gauge of our simulations,
this yields the evolution equation for $W_{lm}$,
\begin {eqnarray}
   \partial_t W_{lm}
   &=&
   \sum_{l'm'} \Bigl\{
     - \langle lm | \v | l'm' \rangle \cdot \D_x W_{l'm'}
\nonumber\\ && \hspace{5em}
     - \left(
         \E \cdot \langle lm | \half [\v,L^2] - \v | l'm' \rangle
         -i \B \cdot \langle lm | \L | l'm' \rangle
       \right) m_\infty^2 \, \Omega_{l'm'} 
   \Bigr\}
\nonumber\\
   &=&
   \sum_{l'm'} \Bigl\{
     - \langle lm | \v | l'm' \rangle \cdot \D_x W_{l'm'}
     + m_\infty^2 \left[ 1+\frac{l(l{+}1)-l'(l'{+}1)}{2} \right] 
       \E\cdot\langle lm | \v | l'm' \rangle \, \Omega_{l'm'}
\nonumber\\ && \hspace{5em} {}
     + i m_\infty^2 \, \B \cdot \langle lm | \L | l'm' \rangle \Omega_{l'm'}
   \Bigr\}
   .
\label {eq:W3}
\end {eqnarray}
The expectation values $\langle lm | \v | l'm' \rangle$ and
$\langle lm | \L | l'm' \rangle$ are simple results from the quantum
mechanics of spin.  The first,
$\langle lm | \v |l'm' \rangle$, is also relevant to the simulations
for isotropic $f_0$,
and explicit formulas may be found in Appendix A of Ref.\ \cite{BMR}.
The $\L$ expectations are
\begin {eqnarray}
  i\,\langle lm | L_x | l'm' \rangle &=& 
  \frac i2 \, \delta_{ll'} \left[
     \delta_{m,m'+1} \sqrt{(l+m)(l-m')}
     + \delta_{m',m+1} \sqrt{(l+m')(l-m)}
   \right] ,
\\
  i\,\langle lm | L_y | l'm' \rangle &=& 
  \frac{1}{2} \, \delta_{ll'} \left[
     \delta_{m,m'+1} \sqrt{(l+m)(l-m')}
     - \delta_{m',m+1} \sqrt{(l+m')(l-m)}
   \right] ,
\\[4pt]
  i\, \langle lm | L_z | l'm' \rangle &=&
  i\, m \,\delta_{ll'} \delta_{mm'} .
\end {eqnarray}

In practice, we found it convenient to implement (\ref{eq:W3}) with
a basis of real functions of $\v$ instead of the usual complex
$Y_{lm}$'s.  So we switched to the basis of
\begin {equation}
   \widetilde Y_{lm} \equiv
   \begin {cases}
      \sqrt2 \, \operatorname{Re} Y_{lm}   ,  &  m > 0; \\
      Y_{lm}                               ,  &  m = 0; \\
      \sqrt2 \, \operatorname{Im} Y_{l|m|} ,  &  m < 0,
   \end {cases}
\end {equation}
with overall normalization again set by (\ref{eq:Ynorm}).


\section {Self energy \boldmath$\Pi$}
\label {app:pi}

In this appendix, we derive the result (\ref{eq:Piperp}) for
the hard-loop transverse gluon self-energy
$\Pi_\perp(\omega,\k)$ when the particle distribution is axi-symmetric
and $\k$ is along the axis of symmetry.
By integration by parts, the spatial part of
Eq.\ (\ref{eq:Pi}) can be
recast into the form \cite{ALM}:
\begin {equation}
   \Pi^{ij}(\omega,\k)
   = e^2 \int_\p \frac{f_0(\p)}{p}
        \left[
          g^{ij}
          - \frac{k^i v^j+k^j v^i}{-\omega+\v\cdot\k-i\epsilon}
          + \frac{(-\omega^2+k^2) v^i v^j}{(-\omega+\v\cdot\k-i\epsilon)^2}
         \right] .
\label {eq:Pi2}
\end {equation}
Take the axis of symmetry to be the $z$ axis.  For $\k$ along that
axis, axial symmetry implies
\begin {equation}
  \Pi_{\perp} \equiv \Pi_{xx} = \Pi_{yy}
        = \half (\delta_{ij} - \hat k_i \hat k_j) \Pi_{ij} .
\label {eq:piperp1}
\end {equation}
Then, using (\ref{eq:Pi2})
and introducing $\eta \equiv {\omega}/{k}$,
\begin {eqnarray}
   \Pi_{\perp}(\omega, k {\bm e}_z)
   &=& e^2 \int_\p \frac{f(\p)}{p}
        \left[
          1 + \frac{(1-\eta^2)[1-(\v\cdot\hat\k)^2]}
                   {2(-\eta+\v\cdot\hat\k)^2}
         \right]
\nonumber\\
   &=&
      \int_{-1}^{+1} \frac{d(\cos\theta)}{2}
          \left[ 1 + \frac{(1-\eta^2)[1-\cos^2\theta]}
                          {2(-\eta+\cos\theta)^2}
          \right]
          {\cal M}(\theta) \,.
\end {eqnarray}
Expanding ${\cal M}$ in spherical harmonics gives
\begin {equation}
   \Pi_{\perp}(\omega, k {\bm e}_z)
   =
      \sum_{l}
      {\cal M}_{l0} \,
      \sqrt{2l{+}1}
      \int_{-1}^{+1} \frac{dx}{2} \>
          \left[ 1 + \frac{(1-\eta^2)(1-x^2)}
                          {2(-\eta+x)^2}
          \right]
          P_l(x).
\end {equation}
Decomposing by partial fractions and using
\begin {equation}
   \int_{-1}^{+1} dx\> \frac{P_l(x)}{-\eta+x}
   = -2 Q_l(\eta) \,,
\end {equation}
and its $\eta$-derivative
\begin {equation}
   \int_{-1}^{+1} dx\> \frac{P_l(x)}{(-\eta+x)^2}
   = -2 Q_l'(\eta)
   = \frac{2(l+1)}{1-\eta^2}
         \left[ Q_{l+1}(\eta) - \eta \, Q_l(\eta) \right]
\end {equation}
produces the final result (\ref{eq:Piperp}).



\end {document}